\documentclass[runningheads]{llncs}
\usepackage{amsmath,amssymb}
\usepackage{mathtools}
\usepackage{tikz}
\usetikzlibrary{decorations.pathreplacing,positioning, arrows.meta,shapes}
\usepackage{tikzsymbols}
\usepackage{ltl}
\usepackage[linesnumbered,lined,commentsnumbered,ruled,noend]{algorithm2e}

\usepackage{todonotes}

\usepackage{graphicx}                   %
\usepackage{hyperref}                   %

\usepackage{color}
\usepackage{colortbl}
\usetikzlibrary{decorations.pathreplacing}
\usetikzlibrary{arrows,petri,backgrounds, positioning, shapes, patterns,calc,automata,intersections,calc}
\usepackage{wrapfig}
\usepackage{caption}
\usepackage{subcaption}
\usepackage{hyperref}
\usepackage{multirow}
\usepackage{scalerel}
\usepackage{cleveref}
\usepackage{array}
\usepackage{comment}
\usepackage{listings}
\usepackage{algorithmic}
\usepackage{booktabs}
\usepackage{moresize}
\usepackage{xcolor}

\renewcommand{\models}{\vDash}

\newcommand{\donotshow}[1]{}

\newtoggle{showmarks}
\toggletrue{showmarks}

\iftoggle{showmarks}{
  \newcommand\daniel[1]{\textcolor{red}{\footnotesize DANIEL: #1}}
  \newcommand\franky[1]{\textcolor{olive}{\footnotesize FRANKY: #1}}
  \newcommand\caroline[1]{\textcolor{orange}{\footnotesize Caroline: #1}}
}{
  \newcommand\daniel[1]{\unskip}
  \newcommand\franky[1]{\unskip}
  \newcommand\caroline[1]{\unskip}
}
\usepackage{fancyvrb}

\colorlet{Gray}{black!75}
\colorlet{Black}{black}

\RecustomVerbatimCommand{\VerbatimInput}{VerbatimInput}%
{fontsize=\scriptsize,
 frame=lines,  %
 framesep=2em, %
 rulecolor=\color{Gray},
 label=\fbox{\color{Black}minimal.txt},
 labelposition=topline,
 commandchars=\|\(\), %
 commentchar=*        %
}

\makeatletter
\def\moverlay{\mathpalette\mov@rlay}
\def\mov@rlay#1#2{\leavevmode\vtop{%
		\baselineskip\z@skip \lineskiplimit-\maxdimen
		\ialign{\hfil$\m@th#1##$\hfil\cr#2\crcr}}}
\newcommand{\charfusion}[3][\mathord]{
	#1{\ifx#1\mathop\vphantom{#2}\fi
		\mathpalette\mov@rlay{#2\cr#3}
	}
	\ifx#1\mathop\expandafter\displaylimits\fi}
\makeatother

\newcolumntype{H}{>{\setbox0=\hbox\bgroup}c<{\egroup}@{}}

\definecolor{TodoRed}{RGB}{225,63,63}
\newcommand{\ttodo}[4]{\ifthenelse{\equal{#1}{inline}}{\todo[inline, author=#2, color=#3,bordercolor=#3,backgroundcolor=#3!25,linecolor=#3]{#4}}{\todo[color=#3,bordercolor=#3,backgroundcolor=#3!25,linecolor=#3]{#2: #4}}}

\definecolor{trace0}{RGB}{179, 76, 18}
\definecolor{trace1}{RGB}{179, 152, 18}
\definecolor{expl0}{RGB}{141, 63, 144}
\definecolor{expl1}{RGB}{236, 77, 216}
\definecolor{highlight}{RGB}{0, 123, 255}

\makeatletter
\newcommand{\printfnsymbol}[1]{%
  \textsuperscript{\@fnsymbol{#1}}%
}
\makeatother

\begin{document}

\title{\texttt{nl2spec}: Interactively Translating Unstructured Natural Language to Temporal Logics with Large Language Models}

\author{Matthias Cosler\inst{2}, Christopher Hahn\inst{1} \and Daniel Mendoza\inst{1} \and Frederik Schmitt\inst{2} \and Caroline Trippel\inst{1}}
\institute{Stanford University, Stanford, CA, USA\\ \email{hahn@cs.stanford.edu, dmendo@stanford.edu, trippel@stanford.edu}
\and
CISPA Helmholtz Center for Information Security, Saarbrücken, Germany \\ \email{matthias.cosler@cispa.de, frederik.schmitt@cispa.de}
}

\authorrunning{Cosler, Hahn, Mendoza, Schmitt, Trippel}
\titlerunning{\texttt{nl2spec}}

\maketitle
\begin{abstract}
A rigorous formalization of desired system requirements is indispensable when performing any verification task.
This often limits the application of verification techniques, as writing formal specifications is an error-prone and time-consuming manual task.
To facilitate this, we present \texttt{nl2spec}, a framework for applying Large Language Models (LLMs) to derive formal specifications (in temporal logics) from unstructured natural language.
In particular, we introduce a new methodology to detect and resolve the inherent ambiguity of system requirements in natural language:
we utilize LLMs to map subformulas of the formalization back to the corresponding natural language fragments of the input.
Users iteratively add, delete, and edit these sub-translations to amend erroneous formalizations, which is easier than manually redrafting the entire formalization.
The framework is agnostic to specific application domains and can be extended to similar specification languages and new neural models.
We perform a user study to obtain a challenging dataset, which we use to run experiments on the quality of translations.
We provide an open-source implementation, including a web-based frontend.
\end{abstract}

\section{Introduction}
A rigorous formalization of desired system requirements is indispensable when performing any verification-related task, such as model checking~\cite{clarke1997model}, synthesis~\cite{church1963application}, or runtime verification~\cite{havelund2001monitoring}.
Writing formal specifications, however, is an error-prone and time-consuming manual task typically reserved for experts in the field.
This paper presents \texttt{nl2spec}, a framework, accompanied by a web-based tool, to facilitate and automate writing formal specifications (in LTL~\cite{pnueli1977temporal} and similar temporal logics).
The core contribution is a new methodology to decompose the natural language input into \emph{sub-translations} by utilizing Large Language Models (LLMs).
The \texttt{nl2spec} framework provides an interface to interactively add, edit, and delete these \emph{sub-translations} instead of attempting to grapple with the entire formalization at once (a feature that is sorely missing in similar work, e.g.,~\cite{fuggittinl2ltl,liu2022lang2ltl}).

\begin{figure}[t]
    \centering
    \includegraphics[width=.95\textwidth]{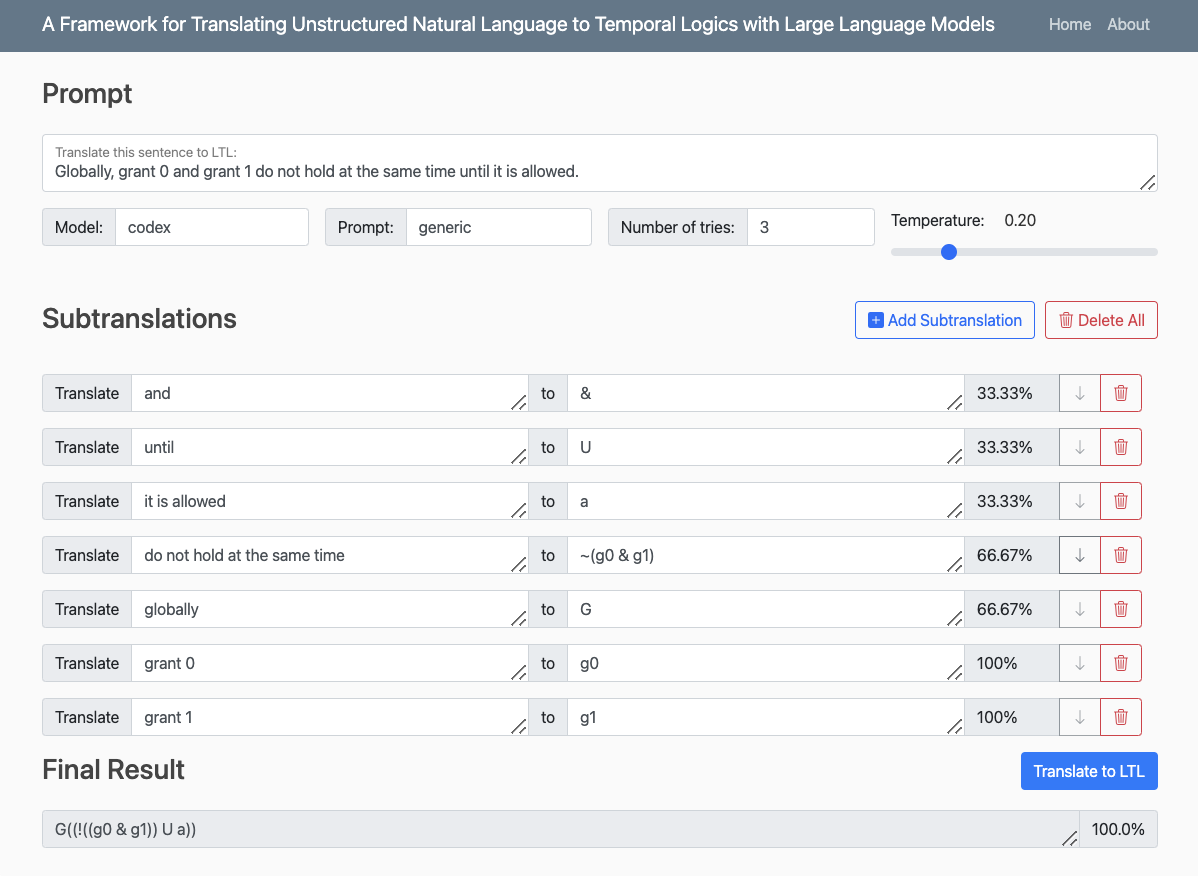}
    \caption{A screenshot of the web-interface for \texttt{nl2spec}.}
    \label{fig:running_example}
\end{figure}

Figure~\ref{fig:running_example} shows the web-based frontend of \texttt{nl2spec}.
As an example, we consider the following system requirement given in natural language: ``Globally, grant 0 and grant 1 do not hold at the same time until it is allowed''.
The tool automatically translates the natural language specification correctly into the LTL formula \texttt{G((!((g0 \& g1)) U a))}.
Additionally, the tool generates sub-translations, such as the pair (``do not hold at the same time'', \texttt{!(g0 \& g1)}), which help in verifying the correctness of the translation.

Consider, however, the following ambiguous example: ``a holds until b holds or always a holds''.
Human supervision is needed to resolve the ambiguity on the operator precedence.
This can be easily achieved with \texttt{nl2spec} by adding or editing a sub-translation using explicit parenthesis (see Section~\ref{sec:experiments} for more details and examples).
To capture such (and other types of) ambiguity in a benchmark data set, we conducted an expert user study specifically asking for challenging translations of natural language sentences to LTL formulas.

The key insight in the design of \texttt{nl2spec} is that the process of translation can be decomposed into many sub-translations automatically via LLMs, and the decomposition into sub-translations allows users to easily resolve ambiguous natural language and erroneous translations through interactively modifying sub-translations.
The central goal of \texttt{nl2spec} is to keep the human supervision minimal and efficient.
To this end, all translations are accompanied by a confidence score,
 alternative suggestions for sub-translations can be displayed and chosen via a drop-down menu, and misleading sub-translations can be deleted before the next loop of the translation.
We evaluate the end-to-end translation accuracy of our proposed methodology on the benchmark data set obtained from our expert user study.

The framework is agnostic to machine learning models and specific application domains.
We will discuss possible parameterizations and inputs of the tool in Section~\ref{sec:framework_overview}.
We discuss our sub-translation methodology in more detail in Section~\ref{sec:interactive_few-shot_prompting} and introduce an interactive few-shot prompting scheme for LLMs to generate them.
We discuss how users can apply this scheme to their respective application domains to increase the quality of the framework's translations.
As proof of concept, we provide additional prompts, including a prompt for STL~\cite{maler2004monitoring} in the appendix\footnote{The tool is available at GitHub: \url{https://github.com/realChrisHahn2/nl2spec}.}.
We evaluate the effectiveness of the tool to resolve erroneous formalizations in Section~\ref{sec:experiments} on a data set obtained from conducting an expert user study. We discuss limitations of the framework and conclude in Section~\ref{sec:conclusion}.

\section{Background \& Related Work}
\label{sec:prelims}

\subsection{Natural Language to Linear-time Temporal Logic}
Linear-time Temporal Logic (LTL)~\cite{pnueli1977temporal} is a temporal logic that forms the basis of many practical specification languages, such as the IEEE property specification language (PSL)~\cite{psl}, Signal Temporal Logic (STL)~\cite{maler2004monitoring}, or System Verilog Assertions (SVA)~\cite{vijayaraghavan2005practical}.
By focusing on the prototype temporal logic LTL, we keep the \texttt{nl2spec} framework extendable to specification languages in specific application domains.
LTL extends propositional logic with temporal modalities \texttt{U} (until) and \texttt{X} (next).
There are several derived operators, such as
$\texttt{F} \varphi \equiv \mathit{true} \texttt{U} \varphi$ and
$\texttt{G} \varphi \equiv \neg \texttt{F} \neg \varphi$.
$\texttt{F} \varphi$ states that $\varphi$ will \emph{eventually} hold in the future and $\texttt{G} \varphi$ states that $\varphi$ holds \emph{globally}.
Operators can be nested: $\texttt{G} \texttt{F} \varphi$, for example, states that $\varphi$ has to occur infinitely often.
LTL specifications describe a systems behavior and its interaction with an environment over time. For example given a process $0$ and a process $1$ and a shared ressource, the formula $\texttt{G} (r_0 \rightarrow \texttt{F} g_0) \wedge \texttt{G} (r_1 \rightarrow \texttt{F} g_1) \wedge \texttt{G} \neg (g_0 \wedge g_1)$ describes that whenever a process requests $(r_i)$ access to a shared ressource it will eventually be granted $(g_i)$. The subformula $\texttt{G} \neg (g_0 \wedge g_1)$ ensures that grants given are mutually exclusive. The formal syntax and semantics of LTL are in Appendix~\ref{app:ltl}.

Early work in translating natural language to temporal logics focused on grammar-based approaches that could handle structured natural language~\cite{konrad2005real,grunske2008specification}. A survey of earlier research before the advent of deep learning is provided in~\cite{brunello2019synthesis}.
Other approaches include an interactive method using SMT solving and semantic parsing~\cite{gavran2020interactive}, or structured temporal aspects in grounded robotics~\cite{wang2020learning} and planning~\cite{patel2019learning}.
Neural networks have only recently being used to translate into temporal logics, e.g., by training a model for STL from scratch~\cite{he2022deepstl}, fine-tuning language models~\cite{hahn2022formal}, or an approach to apply GPT-3~\cite{fuggittinl2ltl,liu2022lang2ltl} in a one-shot fashion, where~\cite{fuggittinl2ltl} output a restricted set of declare templates~\cite{pesic2006declarative} that can be translated to a fragment of LTLf~\cite{de2013linear}.
Translating natural langauge to LTL has especially been of interest to the robotics community (see~\cite{gopalan2018sequence} for an overview), where datasets and application domains are, in contrast to our setting, based on structured natural language.
Independent of relying on structured data, all previous tools lack a detection and interactive resolving of the inerherent ambiguity of natural language, which is the main contribution of our framework.

\subsection{Large Language Models}
\label{sec:large_language_models}

LLMs are large neural networks typically consisting of up to $176$ billion parameters.
They are pre-trained on massive amounts of data, such as ``The Pile''~\cite{gao2020pile}.
Examples of LLMs include the GPT~\cite{radford2018improving} and BERT~\cite{devlin2018bert} model families, open-source models, such as T5~\cite{2020t5} and Bloom~\cite{scao2022bloom}, or commercial models, such as Codex~\cite{chen2021evaluating}.
LLMs are Transformers~\cite{vaswani2017attention}, which is the state of the art neural architecture for natural language proccessing.
Additionally, Transformers have shown remarkable performance when being applied to classical problems in verification (e.g.,~\cite{hahn2020teaching,schmitt2021neural,kreber2021generating,cosler2023repair}), reasoning (e.g., \cite{lewkowycz2022solving,zelikman2022star}), as well as the auto-formalization~\cite{rabe2021towards} of mathematics and formal specifications (e.g.,~\cite{wu2022autoformalization,hahn2022formal,he2022deepstl}).

In language modelling, we model the probability of a sequence of tokens in a text~\cite{shannon1948mathematical}.
The joint probability of tokens in a text is generally modelled as~\cite{scao2022bloom}:
\begin{align*}
    p(x) = p(x_1,\ldots,x_T) = \prod^T_{t=1}p(x_t|x_{<t}) \enspace ,
\end{align*}
where $x$ is the sequence of tokens, $x_t$ represents the $t$-th token, and $x_{<t}$ is the sequence of tokens preceding $x_t$.
We refer tho this as an autoregressive language model that iteratively predicts the probability of the next token.
Neural network approaches to language modelling have superseded classical approaches, such as $n$-grams~\cite{shannon1948mathematical}.
Especially Transformers~\cite{vaswani2017attention} were shown to be the most effective architecture at the time of writing~\cite{radford2018improving,al2019character,kaplan2020scaling} (see Appendix~\ref{app:transformer} for details).

While fine-tuning neural models on a specific translation task remains a valid approach showing also initial success in generalizing to unstructured natural language when translating to LTL~\cite{hahn2022formal}, a common technique to obtain high performance with limited amount of labeled data is so-called ``few-shot prompting''~\cite{brown2020language}.
The language model is presented a natural language description of the task usually accompanied with a few examples that demonstrate the input-output behavior. The framework presented in this paper relies on this technique. We describe the proposed few-shot prompting scheme in detail in Section~\ref{sec:interactive_few-shot_prompting}.

Currently implemented in the framework and used in the expert-user study are Codex and Bloom, which showed the best performance during testing.

\paragraph{Codex.}
Codex~\cite{chen2021evaluating} is a GPT-3 variant that was initially of up to $12$B parameters in size and fine-tuned on code.
The initial version of GPT-3 itself was trained on variations of Common Crawl,\footnote{https://commoncrawl.org/} Webtext-2~\cite{radford2019language}, two internet-based book corpora and Wikipedia~\cite{brown2020language}.
The fine-tuning dataset for the vanilla version Codex was collected in May 2020 from $54$ million public software repositories hosted on GitHub, using $159$GB of training data for fine-tuning.
For our experiments, we used the commercial 2022 version of \texttt{code-davinci-002}, which is likely larger (in the $176$B range\footnote{https://blog.eleuther.ai/gpt3-model-sizes/}) than the vanilla codex models.

\paragraph{Bloom.}
Bloom~\cite{scao2022bloom} is an open-source LLM family available in different sizes of up to $176$B parameters trained on $46$ natural languages and $13$ programming languages. It was trained on the \texttt{ROOTS} corpus~\cite{laurenccon2022bigscience}, a collection of $498$ huggingface~\cite{wolf2020transformers,lhoest2021datasets} datasets consisting of $1.61$ terabytes of text.
For our experiments, we used the $176$B version running on the huggingface inference API~\footnote{https://huggingface.co/inference-api}.

\section{The \texttt{nl2spec} Framework}
\label{sec:framework_overview}

\begin{figure}[t]
    \centering
    \def\svgwidth{0.76\textwidth}\ssmall
    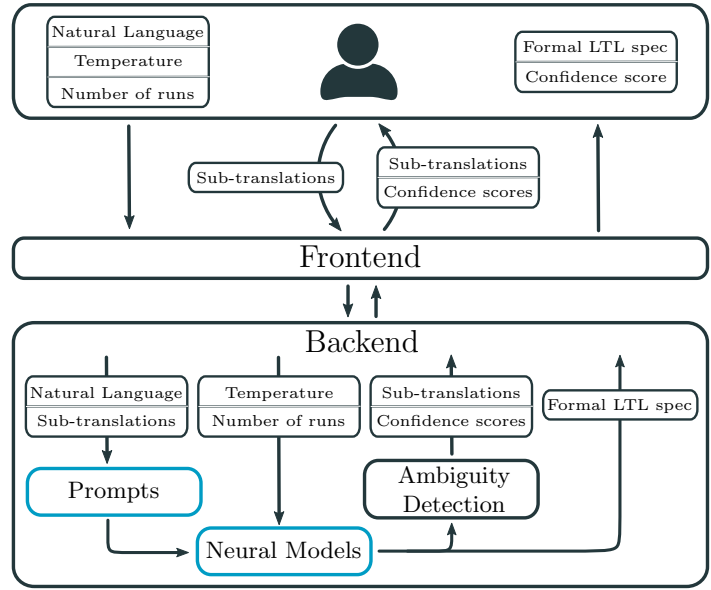
    \caption{Overview of the \texttt{nl2spec} framework with a human-in-the-loop: highlighted areas indicate parts of the framework that are effortlessly extendable.}
    \label{fig:overview}
\end{figure}

\subsection{Overview}

The framework follows a standard frontend-backend implementation.
Figure~\ref{fig:overview} shows an overview of the implementation of \texttt{nl2spec}.
Parts of the framework that can be extended for further research or usage in practice are highlighted.
The framework is implemented in Python 3 and flask~\cite{vyshnavi2019efficient}, a lightweight WSGI web application framework. For the experiments in this paper, we use the OpenAI library and huggingface (transformer) library~\cite{wolf2019huggingface}.
We parse the LTL output formulas with a standard LTL parser~\cite{Fuggitti_2019}.
The tool can either be run as a command line tool, or with the web-based frontend.

The frontend handles the interaction with a human-in-the-loop.
The interface is structured in three views: the ``Prompt'', ``Sub-translations'', and ``Final Result'' view (see Fig.~\ref{fig:running_example}).
The tool takes a natural language sentence, optional sub-translations, the model temperature, and number of runs as input.
It provides sub-translations, a confidence score, alternative subtranslations and the final formalization as output.
The frontend then allows for interactively selecting, editing, deleting, or adding sub-translations.
The backend implements the handling of the underlying neural models, the generation of the prompt, and the ambiguity resolving, i.e., computing the confidence score including alternative subtranslations and the interactive few-shot prompting algorithm (cf. Section~\ref{sec:interactive_few-shot_prompting}).
The framework is designed to have an easy interface to implement new models and write domain-specific prompts.
The prompt is a .txt file that can be adjusted to specific domains to increase the quality of translations (see Appendix~\ref{app:prompts}). To apply the methodology of the framework, however, the prompt needs to follow our interactive prompting scheme, which we introduce in the next section.

\begin{figure}[t]
\begin{Verbatim}
Translate the following natural language sentences into an LTL formula and explain your
translation step by step. Remember that X means "next", U means "until", G means
"globally", F means "finally", which means GF means "infinitely often". The formula
should only contain atomic propositions or operators |, &, ~, ->, <->, X, U, G, F.
Natural Language: Globally if a holds then c is true until b. Given translations: {}
Explanation: "a holds" from the input translates to the atomic proposition a.
"c is true until b" from the input translates to the subformula c U b. "if x then y"
translates to an implication x -> y, so "if a holds then c is true until b" translates
to an implication a -> c U b. "Globally" from the input translates to the temporal
operator G. Explanation dictionary: {"a holds" : "a", "c is true until b" : "c U b",
"if a holds then c is true until b" : "a -> c U b", "Globally" : "G"} So the final
LTL translation is G (a -> (c U b)).FINISH Natural Language: Every request r is
eventually followed by a grant g. Given translations: {} Explanation: "Request r"
from the input translates to the atomic proposition r and "grant g" translates to the
atomic proposition g. "every" means at every point in time, i.e., globally, "never"
means at no point in time, and "eventually" translates to the temporal operator F.
"followed by" is the natural language representation of an implication. Explanation
dictionary: {"Request r" : "r", "grant g" : "g", "every" : "G", "eventually": "F",
"followed by" : "->"} So the final LTL translation is G (r -> F g).FINISH
\end{Verbatim}
\caption{Prompt with minimal domain knowledge of LTL.}
\label{fig:generic_prompt}
\end{figure}

\subsection{Interactive Few-shot Prompting}
\label{sec:interactive_few-shot_prompting}
The core of the methodology is the decomposition of the natural language input into sub-translations. We introduce an interactive prompting scheme that generates sub-translations using the underlying neural model and leverages the sub-translations to produce the final translation.
Algorithm~\ref{alg:algorithm} depicts an high-level overview of the interactive loop.
The main idea is to give a human-in-the-loop the options to add, edit, or delete sub-translations and feed them back into the language models as ``Given translations'' in the prompt (see Fig.~\ref{fig:generic_prompt}).
After querying a language model $M$ with this prompt $F$, model specific parameters $P$ and the interactive prompt that is computed in the loop, the model generates a natural language explanation, a dictionary of subtranslations, and the final translation.
The confidence scores are computed as votes over multiple queries to $M$, where the (sub) translation with the highest consensus score is displayed; alternative translations can be displayed and chosen interactively by clicking on the downarrow.

Figure~\ref{fig:generic_prompt} shows a generic prompt, that illustrates our methodology.
The prompting scheme consists of three parts.
The specification language specific part (lines $1-4$), the fewshot examples (lines $5-19$), and the interactive prompt including the natural language and sub-translation inputs (not displayed, given as input).
The specification language specific part includes prompt-engineering tricks taken from ``chain-of-thought'' generation to elicit reasoning from large language models~\cite{wei2022chain}.
The key of \texttt{nl2spec}, however, is the setup of the few-shot examples.
This minimal prompt consists of two few-shot examples (lines $5-12$ and $12-19$).
The end of an example is indicated by the ``FINISH'' token, which is the stop token for the machine learning models.
A few-shot example in \texttt{nl2spec} consists of the natural language input (line $5$), a dictionary of given translations, i.e., the sub-translations (line $5$), an explanation of the translation in natural language (line $6-10$), an explanation dictionary, summarizing the sub-translations, and finally, the final LTL formula.

This prompting scheme elicits sub-translations from the model, which serve as a fine-grained explanation of the formalization.
Note that sub-translations provided in the prompt are neither unique nor exhaustive, but provide the context for the language model to generate the correct formalization.

\begin{algorithm}[t]
    \caption{Interactive Few-shot Prompting Algorithm}
    \label{alg:algorithm}
    \textbf{Input}: Natural language $S$, Few-shot prompt $F$, set of given sub-translations $(s,\varphi)$, and language model $M$\\
    \textbf{Interactions:} set of sub-translations $(s,\varphi)$, confidence scores $C$\\
    \textbf{Set of Model specific parameter $P$}: e.g., model-temperature $t$, number of runs $r$\\
    \textbf{Output}: LTL formula $\psi$ that formalizes $S$
    \begin{algorithmic}[1] %
        \STATE $\psi,~(s,\varphi)~,C$ = empty
        \WHILE{user not approves LTL formula $\psi$}
        \STATE interactive\_prompt = \texttt{compute\_prompt}(S, F, $(s,\varphi)$)\\
        \STATE $\psi,~(s,\varphi)~,C$ = \texttt{query}(M, P, interactive\_prompt)\\
        \STATE $(s,\varphi)$ = \texttt{user\_interaction}($(s,\varphi)~,C$)
        \ENDWHILE
        \STATE \textbf{return} $\psi$
    \end{algorithmic}
\end{algorithm}

\section{Evaluation}
\label{sec:experiments}

In this section, we evaluate our framework and prompting methodology on a data set obtained by conducting an expert user study.
To show the general applicability of this framework, we use the \texttt{minimal} prompt that includes only minimal domain knowledge of the specification language (see Figure~\ref{fig:generic_prompt}).
This prompt has intentionally been written \emph{before} conducting the expert user study.
We limited the few-shot examples to two and even provided no few-shot example that includes ``given translations''.
We use the minimal prompt to focus the evaluation on the effectiveness of our interactive sub-translation refinement methodology in resolving ambiguity and fixing erroneous translations.
In practice, one would like to replace this minimal prompt with domain-specific examples that capture the underlying distribution as closely as possible. As a proof of concept, we elaborate on this in Appendix~\ref{app:prompts}.

\subsection{Study Setup}
To obtain a benchmark dataset of \emph{unstructured} natural language and their formalizations into LTL, we asked five experts in the field to provide examples that the experts thought are challenging for a neural translation approach.
Unlike existing datasets that follow strict grammatical and syntatical structure, we posed no such restrictions on the study participants.
Each natural language specification was restricted to one sentence and to five atomic propositions $a,b,c,d,e$.
Note that \texttt{nl2spec} is not restricted to a specific set of atomic propositions (cf. Figure~\ref{fig:running_example}). Which variable scheme to use can be specified as an initial sub-translation. We elaborate on this in Appendix~\ref{app:variables}.
To ensure unique instances, the experts worked in a shared document, resulting in $36$ benchmark instances.
We provide three randomly drawn examples for the interested reader:

{\centering
\resizebox{\linewidth}{!}{
\begin{tabular}{c | c}
    natural language S & LTL specification $\psi$ \\
    \midrule
    If b holds then, in the next step, c holds until a holds or always c holds. & \texttt{b -> X ((c U a) || G c)} \\
    \midrule
    If b holds at some point, a has to hold somewhere beforehand. & \texttt{(F b) -> (!b U (a \& !b))} \\
    \midrule
    One of the following aps will hold at all instances: a,b,c. & \texttt{G( a | b | c)} \\
    \bottomrule
\end{tabular}
}
}

The poor performance of existing methods (cf. Table~\ref{table:initial}) exemplify the difficulty of this data set.

\subsection{Results}

We evaluated our approach using the \texttt{minimal} prompt (if not otherwise stated), with number of runs set to three and with a temperature of $0.2$.

\vspace{1ex}

\noindent
\emph{Quality of Initial Translation.}
We analyze the quality of \emph{initial} translations, i.e., translations obtained \emph{before} any human interaction.
This experiment demonstrates that the initial translations are of high quality, which is important to ensure an efficient workflow.
We compared our approach to fine-tuning language models on structured data~\cite{hahn2022formal} and to an approach using GPT-3 or Rasa~\cite{bocklisch2017rasa} to translate natural language into a restricted set of declare patterns~\cite{fuggittinl2ltl} (which could not handle most of the instances in the benchmark data set, even when replacing the atomic propositions with their used entities).
The results of evaluating the accuracy of the initial translations on our benchmark expert set is shown in Table~\ref{table:initial}.

At the time of writing, using Codex in the backend outperforms Bloom on this task, by correctly translating $44.4\%$ of the instances using the minimal prompt.
We only count an instance as correctly translated if it matches the intended meaning of the expert, no alternative translation to ambiguous input was accepted.
Additionally to the experiments using the minimal prompt, we conducted experiments on an augmented prompt with in-distribution examples after the user study was conducted by randomly drawing four examples from the expert data set ($3$ of the examples haven't been solved before, see Appendix~\ref{app:prompts}).
With this in-distribution prompt (ID), the tool translates $21$ instances (with the four drawn examples remaining in the set), i.e., $58.3\%$ correctly.

\begin{table*}[b]
\caption{Initial Translation accuracy on the benchmark data set, where B stands for Bloom and C stands for Codex.}
\label{table:initial}
{\centering
\resizebox{\linewidth}{!}{
\begin{tabular}{c | c | c | c | c | c}
    \toprule
     nl2ltl\cite{fuggittinl2ltl} & T-5~\cite{hahn2022formal} & \texttt{nl2spec}+B & \texttt{nl2spec}+C  & \texttt{nl2spec}+C & \texttt{nl2spec}+C \\
     \texttt{rasa} & \texttt{fine-tuned} & \texttt{initial} &  \texttt{initial} & \texttt{initial+ID} & \texttt{interactive} \\
    \midrule
    1/36 (2.7\%) & 2/36 (5.5\%) & 5/36 (13.8\%) & 16/36 (44.4\%) & 21/36 (58.3\%) & 31/36 (86.11\%) \\
    \bottomrule
\end{tabular}
}
}
\end{table*}

This experiment shows 1) that the initial translation quality is high and can handle unstructured natural language better than previous approaches and 2) that drawing the few-shot examples in distribution only slightly increased translation quality for this data set; making the key contributions of \texttt{nl2spec}, i.e., ambiguity detection and effortless debugging of erroneous formalizations, valuable.
Since \texttt{nl2spec} is agnostic to the underlying machine learning models, we expect an even better performance in the future with more fine-tuned models.

\vspace{1ex}

\noindent
\emph{Teacher-Student Experiment.}
In this experiment, we generate an initial set of sub-translations with Codex as the underlying neural model. We then ran the tool with Bloom as a backend, taking these sub-translations as input.
There were $11$ instances that Codex could solve initially that Bloom was unable to solve.
On these instances, Bloom was able to solve $4$ more instances, i.e., $36.4\%$ with sub-translations provided by Codex.
The four instances that Bloom was able to solve with the help of Codex were: ``It is never the case that a and b hold at the same time.'', ``Whenever a is enabled, b is enabled three steps later.'', ``If it is the case that every a is eventually followed by a b, then c needs to holds infinitely often.'', and ``One of the following aps will hold at all instances: a,b,c''.
This demonstrates that our sub-translation methodology is a valid appraoch: improving the quality of the sub-translations indeed has a positive effect on the quality of the final formalization. This even holds true when using underperforming neural network models.
Note that no supervision by a human was needed in this experiment to improve the formalization quality.

\vspace{1ex}

\noindent
\emph{Ambiguity Detection.}
Out of the $36$ instances in the benchmark set, at least $9$ of the instances contain ambiguous natural language.
We especially observed two classes of ambiguity: 1) ambiguity due to the limits of natural language, e.g., operator precedence,
and 2) ambiguity in the semantics of natural language;
\texttt{nl2spec} can help in resolving both types of ambiguity.

An example for the first type of ambiguity from our dataset is the example mentioned in the introduction:
``a holds until b holds or always a holds'', which the expert translated into \texttt{(a U b) | G a}.
Running the tool, however, translated this example into \texttt{(a U (b | G(a)))} (see Appendix~\ref{app:ambiguity_resolving}).
By editting the sub-translation of ``a holds until b holds'' to \texttt{(a U b)} through adding explicit parenthesis, the tool translates as intended.
An example for the second type of ambiguity is the following instance from our data set:
``Whenever a holds, b must hold in the next two steps.''
The intended meaning of the expert was \texttt{G (a -> (b | X b))}, whereas the tool translated this sentence into \texttt{G((a -> X(X(b))))}.
After changing the sub-translation of ``b must hold in the next two steps'' to \texttt{b | X b}, the tool translates the input as intended (see Appendix~\ref{app:ambiguity_resolving}).

\vspace{1ex}

\noindent
\emph{Fixing Erroneous Translation.}
With the inherent ambiguity of natural language and the unstructured nature of the input, the tool's translation cannot be expected to be always correct in the first try.
Verifying and debugging sub-translations, however, is significantly easier than redrafting the complete formula from scratch.
Twenty instances of the data set were not correctly translated in an initial attempt using Codex and the minimal prompt in the backend (see Table~\ref{table:initial}).
We were able to extract correct translations for $15$ instances by performing at most three translation loops (i.e., adding, editing, and removing sub-translations),
We were able to get correct results by performing $1.86$ translation loops on average.
For example, consider the instance, ``whenever a holds, b holds as well'', which the tool mistakenly translated to \texttt{G(a \& b)}. By fixing the sub-translation ``b holds as well'' to the formula fragment \texttt{-> b}, the sentence is translated as intended (see Appendix~\ref{app:fixing}).
Only the remaining five instances that contain highly complex natural language requirements, such as, ``once a happened, b won't happen again'' were need to be translated by hand (see Appendix~\ref{app:remaining_instances}).

In total, we correctly translated $31$ out of $36$ instances, i.e., $86.11\%$ using the \texttt{nl2spec} sub-translation methodology by performing only $1.4$ translation loops on average (see Table~\ref{table:initial}).

\section{Conclusion}
\label{sec:conclusion}

We presented \texttt{nl2spec}, a framework for translating unstructured natural language to temporal logics.
A limitation of this approach is its reliance on computational ressources at inference time.
This is a general limitation when applying deep learning techniques.
Both, commercial and open-source models, however, provide easily accessible APIs to their models.
Additionally, the quality of initial translations might be influenced by the amount of training data on logics, code, or math that the underlying neural models have seen during pre-training.

At the core of \texttt{nl2spec} lies a methodology to decompose the natural language input into sub-translations, which are mappings of formula fragments to relevant parts of the natural language input.
We introduced an interactive prompting scheme that queries Large Language Models (LLMs) for sub-translations, and implemented an interface for users to interactively add, edit, and delete the sub-translations, which eschews users from manually redrafting the entire formalization to fix erroneous translations.
We conducted a user study, showing that \texttt{nl2spec} can be efficiently used to interactively formalize unstructured and ambigous natural language.

\section*{Acknowledgements}
We thank OpenAI for providing academic access to Codex and Clark Barrett for helpful feedback on an earlier version of the tool.

\bibliographystyle{splncs04}
\bibliography{bibliography}

\begin{thebibliography}{10}
\providecommand{\url}[1]{\texttt{#1}}
\providecommand{\urlprefix}{URL }
\providecommand{\doi}[1]{https://doi.org/#1}

\bibitem{jarvis}
{J.A.R.V.I.S.} {TSL}/{TLSF} benchmark suite (2021),
  \url{https://github.com/SYNTCOMP/benchmarks/tree/master/tlsf/tsl_smart_home_jarvis}

\bibitem{al2019character}
Al-Rfou, R., Choe, D., Constant, N., Guo, M., Jones, L.: Character-level
  language modeling with deeper self-attention. In: Proceedings of the AAAI
  conference on artificial intelligence. vol.~33, pp. 3159--3166 (2019)

\bibitem{bocklisch2017rasa}
Bocklisch, T., Faulkner, J., Pawlowski, N., Nichol, A.: Rasa: Open source
  language understanding and dialogue management. arXiv preprint
  arXiv:1712.05181  (2017)

\bibitem{brown2020language}
Brown, T., Mann, B., Ryder, N., Subbiah, M., Kaplan, J.D., Dhariwal, P.,
  Neelakantan, A., Shyam, P., Sastry, G., Askell, A., et~al.: Language models
  are few-shot learners. Advances in neural information processing systems
  \textbf{33},  1877--1901 (2020)

\bibitem{brunello2019synthesis}
Brunello, A., Montanari, A., Reynolds, M.: Synthesis of ltl formulas from
  natural language texts: State of the art and research directions. In: 26th
  International Symposium on Temporal Representation and Reasoning (TIME 2019).
  Schloss Dagstuhl-Leibniz-Zentrum fuer Informatik (2019)

\bibitem{chen2021evaluating}
Chen, M., Tworek, J., Jun, H., Yuan, Q., Pinto, H.P.d.O., Kaplan, J., Edwards,
  H., Burda, Y., Joseph, N., Brockman, G., et~al.: Evaluating large language
  models trained on code. arXiv preprint arXiv:2107.03374  (2021)

\bibitem{church1963application}
Church, A.: Application of recursive arithmetic to the problem of circuit
  synthesis. Journal of Symbolic Logic  \textbf{28}(4) (1963)

\bibitem{clarke1997model}
Clarke, E.M.: Model checking. In: International Conference on Foundations of
  Software Technology and Theoretical Computer Science. pp. 54--56. Springer
  (1997)

\bibitem{cosler2023repair}
Cosler, M., Schmitt, F., Hahn, C., Finkbeiner, B.: Iterative circuit repair
  against formal specifications. In: International Conference on Learning
  Representations (to appear) (2023)

\bibitem{de2013linear}
De~Giacomo, G., Vardi, M.Y.: Linear temporal logic and linear dynamic logic on
  finite traces. In: IJCAI'13 Proceedings of the Twenty-Third international
  joint conference on Artificial Intelligence. pp. 854--860. Association for
  Computing Machinery (2013)

\bibitem{devlin2018bert}
Devlin, J., Chang, M.W., Lee, K., Toutanova, K.: Bert: Pre-training of deep
  bidirectional transformers for language understanding. arXiv preprint
  arXiv:1810.04805  (2018)

\bibitem{donze2013signal}
Donz{\'e}, A.: On signal temporal logic. In: Runtime Verification: 4th
  International Conference, RV 2013, Rennes, France, September 24-27, 2013.
  Proceedings 4. pp. 382--383. Springer (2013)

\bibitem{Fuggitti_2019}
Fuggitti, F.: LTLf2DFA. Zenodo (Mar 2019). \doi{10.5281/ZENODO.3888410},
  \url{https://zenodo.org/record/3888410}

\bibitem{fuggittinl2ltl}
Fuggitti, F., Chakraborti, T.: Nl2ltl--a python package for converting natural
  language (nl) instructions to linear temporal logic (ltl) formulas

\bibitem{gao2020pile}
Gao, L., Biderman, S., Black, S., Golding, L., Hoppe, T., Foster, C., Phang,
  J., He, H., Thite, A., Nabeshima, N., et~al.: The pile: An 800gb dataset of
  diverse text for language modeling. arXiv preprint arXiv:2101.00027  (2020)

\bibitem{gavran2020interactive}
Gavran, I., Darulova, E., Majumdar, R.: Interactive synthesis of temporal
  specifications from examples and natural language. Proceedings of the ACM on
  Programming Languages  \textbf{4}(OOPSLA),  1--26 (2020)

\bibitem{gopalan2018sequence}
Gopalan, N., Arumugam, D., Wong, L.L., Tellex, S.: Sequence-to-sequence
  language grounding of non-markovian task specifications. In: Robotics:
  Science and Systems. vol.~2018 (2018)

\bibitem{grunske2008specification}
Grunske, L.: Specification patterns for probabilistic quality properties. In:
  2008 ACM/IEEE 30th International Conference on Software Engineering. pp.
  31--40. IEEE (2008)

\bibitem{hahn2020teaching}
Hahn, C., Schmitt, F., Kreber, J.U., Rabe, M.N., Finkbeiner, B.: Teaching
  temporal logics to neural networks. In: International Conference on Learning
  Representations (2021)

\bibitem{hahn2022formal}
Hahn, C., Schmitt, F., Tillman, J.J., Metzger, N., Siber, J., Finkbeiner, B.:
  Formal specifications from natural language. arXiv preprint arXiv:2206.01962
  (2022)

\bibitem{havelund2001monitoring}
Havelund, K., Ro{\c{s}}u, G.: Monitoring java programs with java pathexplorer.
  Electronic Notes in Theoretical Computer Science  \textbf{55}(2),  200--217
  (2001)

\bibitem{he2022deepstl}
He, J., Bartocci, E., Ni{\v{c}}kovi{\'c}, D., Isakovic, H., Grosu, R.: Deepstl:
  from english requirements to signal temporal logic. In: Proceedings of the
  44th International Conference on Software Engineering. pp. 610--622 (2022)

\bibitem{psl}
IEEE-Commission, et~al.: {IEEE} standard for property specification language
  ({PSL}). IEEE Std 1850-2005  (2005)

\bibitem{kaplan2020scaling}
Kaplan, J., McCandlish, S., Henighan, T., Brown, T.B., Chess, B., Child, R.,
  Gray, S., Radford, A., Wu, J., Amodei, D.: Scaling laws for neural language
  models. arXiv preprint arXiv:2001.08361  (2020)

\bibitem{konrad2005real}
Konrad, S., Cheng, B.H.: Real-time specification patterns. In: Proceedings of
  the 27th international conference on Software engineering. pp. 372--381
  (2005)

\bibitem{kreber2021generating}
Kreber, J.U., Hahn, C.: Generating symbolic reasoning problems with transformer
  gans. arXiv preprint arXiv:2110.10054  (2021)

\bibitem{laurenccon2022bigscience}
Lauren{\c{c}}on, H., Saulnier, L., Wang, T., Akiki, C., del Moral, A.V.,
  Le~Scao, T., Von~Werra, L., Mou, C., Ponferrada, E.G., Nguyen, H., et~al.:
  The bigscience roots corpus: A 1.6 tb composite multilingual dataset. In:
  Thirty-sixth Conference on Neural Information Processing Systems Datasets and
  Benchmarks Track (2022)

\bibitem{lewkowycz2022solving}
Lewkowycz, A., Andreassen, A., Dohan, D., Dyer, E., Michalewski, H., Ramasesh,
  V., Slone, A., Anil, C., Schlag, I., Gutman-Solo, T., et~al.: Solving
  quantitative reasoning problems with language models. arXiv preprint
  arXiv:2206.14858  (2022)

\bibitem{lhoest2021datasets}
Lhoest, Q., del Moral, A.V., Jernite, Y., Thakur, A., von Platen, P., Patil,
  S., Chaumond, J., Drame, M., Plu, J., Tunstall, L., et~al.: Datasets: A
  community library for natural language processing. arXiv preprint
  arXiv:2109.02846  (2021)

\bibitem{liu2022lang2ltl}
Liu, J.X., Yang, Z., Schornstein, B., Liang, S., Idrees, I., Tellex, S., Shah,
  A.: Lang2ltl: Translating natural language commands to temporal specification
  with large language models. In: Workshop on Language and Robotics at CoRL
  2022

\bibitem{maler2004monitoring}
Maler, O., Nickovic, D.: Monitoring temporal properties of continuous signals.
  In: Formal Techniques, Modelling and Analysis of Timed and Fault-Tolerant
  Systems: Joint International Conferences on Formal Modeling and Analysis of
  Timed Systmes, FORMATS 2004, and Formal Techniques in Real-Time and
  Fault-Tolerant Systems, FTRTFT 2004, Grenoble, France, September 22-24, 2004.
  Proceedings. pp. 152--166. Springer (2004)

\bibitem{mikolov2013efficient}
Mikolov, T., Chen, K., Corrado, G., Dean, J.: Efficient estimation of word
  representations in vector space. arXiv preprint arXiv:1301.3781  (2013)

\bibitem{patel2019learning}
Patel, R., Pavlick, R., Tellex, S.: Learning to ground language to temporal
  logical form. NAACL (2019)

\bibitem{pesic2006declarative}
Pesic, M., Van~der Aalst, W.M.: A declarative approach for flexible business
  processes management. In: Business Process Management Workshops: BPM 2006
  International Workshops, BPD, BPI, ENEI, GPWW, DPM, semantics4ws, Vienna,
  Austria, September 4-7, 2006. Proceedings 4. pp. 169--180. Springer (2006)

\bibitem{pnueli1977temporal}
Pnueli, A.: The temporal logic of programs. In: 18th Annual Symposium on
  Foundations of Computer Science (sfcs 1977). pp. 46--57. ieee (1977)

\bibitem{rabe2021towards}
Rabe, M.N., Szegedy, C.: Towards the automatic mathematician. In: International
  Conference on Automated Deduction. pp. 25--37. Springer, Cham (2021)

\bibitem{radford2018improving}
Radford, A., Narasimhan, K., Salimans, T., Sutskever, I., et~al.: Improving
  language understanding by generative pre-training  (2018)

\bibitem{radford2019language}
Radford, A., Wu, J., Child, R., Luan, D., Amodei, D., Sutskever, I., et~al.:
  Language models are unsupervised multitask learners. OpenAI blog
  \textbf{1}(8), ~9 (2019)

\bibitem{2020t5}
Raffel, C., Shazeer, N., Roberts, A., Lee, K., Narang, S., Matena, M., Zhou,
  Y., Li, W., Liu, P.J.: Exploring the limits of transfer learning with a
  unified text-to-text transformer. Journal of Machine Learning Research
  \textbf{21}(140),  1--67 (2020), \url{http://jmlr.org/papers/v21/20-074.html}

\bibitem{scao2022bloom}
Scao, T.L., Fan, A., Akiki, C., Pavlick, E., Ili{\'c}, S., Hesslow, D.,
  Castagn{\'e}, R., Luccioni, A.S., Yvon, F., Gall{\'e}, M., et~al.: Bloom: A
  176b-parameter open-access multilingual language model. arXiv preprint
  arXiv:2211.05100  (2022)

\bibitem{schmitt2021neural}
Schmitt, F., Hahn, C., Rabe, M.N., Finkbeiner, B.: Neural circuit synthesis
  from specification patterns. Advances in Neural Information Processing
  Systems  \textbf{34},  15408--15420 (2021)

\bibitem{shannon1948mathematical}
Shannon, C.E.: A mathematical theory of communication. The Bell system
  technical journal  \textbf{27}(3),  379--423 (1948)

\bibitem{vaswani2017attention}
Vaswani, A., Shazeer, N., Parmar, N., Uszkoreit, J., Jones, L., Gomez, A.N.,
  Kaiser, {\L}., Polosukhin, I.: Attention is all you need. Advances in neural
  information processing systems  \textbf{30} (2017)

\bibitem{vijayaraghavan2005practical}
Vijayaraghavan, S., Ramanathan, M.: A practical guide for SystemVerilog
  assertions. Springer Science \& Business Media (2005)

\bibitem{vyshnavi2019efficient}
Vyshnavi, V.R., Malik, A.: Efficient way of web development using python and
  flask. Int. J. Recent Res. Asp  \textbf{6}(2),  16--19 (2019)

\bibitem{wang2020learning}
Wang, C., Ross, C., Kuo, Y.L., Katz, B., Barbu, A.: Learning a natural-language
  to ltl executable semantic parser for grounded robotics. arXiv preprint
  arXiv:2008.03277  (2020)

\bibitem{wei2022chain}
Wei, J., Wang, X., Schuurmans, D., Bosma, M., Chi, E., Le, Q., Zhou, D.: Chain
  of thought prompting elicits reasoning in large language models. arXiv
  preprint arXiv:2201.11903  (2022)

\bibitem{wolf2019huggingface}
Wolf, T., Debut, L., Sanh, V., Chaumond, J., Delangue, C., Moi, A., Cistac, P.,
  Rault, T., Louf, R., Funtowicz, M., et~al.: Huggingface's transformers:
  State-of-the-art natural language processing. arXiv preprint arXiv:1910.03771
   (2019)

\bibitem{wolf2020transformers}
Wolf, T., Debut, L., Sanh, V., Chaumond, J., Delangue, C., Moi, A., Cistac, P.,
  Rault, T., Louf, R., Funtowicz, M., et~al.: Transformers: State-of-the-art
  natural language processing. In: Proceedings of the 2020 conference on
  empirical methods in natural language processing: system demonstrations. pp.
  38--45 (2020)

\bibitem{wu2022autoformalization}
Wu, Y., Jiang, A.Q., Li, W., Rabe, M.N., Staats, C., Jamnik, M., Szegedy, C.:
  Autoformalization with large language models. arXiv preprint arXiv:2205.12615
   (2022)

\bibitem{zelikman2022star}
Zelikman, E., Wu, Y., Goodman, N.D.: Star: Bootstrapping reasoning with
  reasoning. arXiv preprint arXiv:2203.14465  (2022)

\end{thebibliography}

\appendix
\newpage

\section{Linear-time Temporal Logic (LTL)}
\label{app:ltl}

In this section, we provide the formal syntax and semantics of Linear-time Temporal Logic (LTL) to an interested reader.
Formally, the syntax of LTL is as follows:
\begin{align*}
\varphi~ \Coloneqq~p~~|~~\neg \varphi~~|~~\varphi \wedge \varphi~~|~~\LTLnext \varphi~~|~~\varphi\, \LTLuntil \varphi \enspace ,
\end{align*}
where $p \in \mathit{AP}$ is an atomic proposition.
We define the set of traces $\mathit{TR} \coloneqq (2^\mathit{AP})^\omega$.
We use the following notation to manipulate traces:
Let $t \in \mathit{TR}$ be a trace and $i \in \mathbb{N}$ be a natural number. 
With $t[i]$ we denote the set of propositions at $i$-th position of $t$.
Therefore, $t[0]$ represents the starting element of the trace.
Let $j \in \mathbb{N}$ and $j \geq i$.
Then $t[i,j]$ denotes the sequence $t[i]~t[i+1]\ldots t[j-1]~t[j]$ and $t[i, \infty]$ denotes the infinite suffix of $t$ starting at position~$i$.

Let $p \in \mathit{AP}$ and $t \in \mathit{TR}$.
The semantics of an LTL formula is defined as the smallest relation $\models$ that satisfies the following conditions:
\begin{align*}
& t \models p &&~\text{iff} \hspace{2.5ex} p \in t[0] \\
& t \models \neg \varphi &&~\text{iff} \hspace{2.5ex} t \not \models \varphi \\
& t \models \varphi_1 \wedge \varphi_2 &&~\text{iff} \hspace{2.5ex} t \models \varphi_1~\text{and}~t \models \varphi_2 \\
& t \models \LTLnext \varphi &&~\text{iff} \hspace{2.5ex} t [1,\infty] \models \varphi \\
& t \models \varphi_1 \LTLuntil \varphi_2 &&~\text{iff}\hspace{2.5ex} \text{there exists}~i \geq 0 : t[i,\infty] \models \varphi_2 \\
& &&\hspace{5.3ex} \text{and for all}~0 \leq j < i:~t[j,\infty] \models \varphi_1 \enspace .
\end{align*}

\section{Transformer Architecture}
\label{app:transformer}
The underlying neural network of an LLM is based on variations of the Transformer architecture.
A vanilla Transformer follows a basic encoder-decoder structure.
The encoder constructs a hidden embedding $z_i$ for each input embedding $x_i$ of the input sequence $x = (x0, \ldots, x_n)$ in one go.
An embedding is a mapping from plain input, for example words or characters, to a high dimensional vector, for which learning
algorithms and toolkits exists, e.g., word2vec~\cite{mikolov2013efficient}.
Given the encoders output $z = (z_0, \ldots, z_k)$, the decoder generates a sequence of output embeddings $y = (y_0, \ldots , y_m)$ autoregressively.
Since the transformer architecture contains no recurrence nor any convolution, a positional encoding is added to the input and output embeddings that allows to distinguish between different orderings.
Different LLMs use variations of the Transformer architecture, such as decoder-only models in the GPT model family~\cite{radford2018improving}, or encoder-only models in the BERT model family~\cite{devlin2018bert}.

Instead of maintaining a hidden state, e.g. in a recurrent neural network architecture, the self-attention mechanism allows the neural network to incorporate the hidden embedding of other important input elements into the hidden embedding of the current element under consideration.
For each input embedding $x_i$, we compute 1) a query vector $q_i$, 2) a key vector $k_i$, and 3) a value vector $v_i$ by multiplying $x_i$ with weight matrices $W_k$, $W_v$, and $W_q$, which are learned during the training process.
The main idea of the self-attention mechanism is to compute a score for each
pair $(x_i, x_j)$ representing which positions in the sequence should be considered the
most when computing the embedding of $x_i$.
The embeddings can be calculated all at once using matrix operations~\cite{vaswani2017attention}. Let
$Q, K, V$ be the matrices obtained by multiplying the input vector X consisting of all
$x_i$ with the weight matrices $W_k$, $W_v$, and $W_q$:
\begin{align*}
\mathit{Attention}(Q, K, V) = \mathit{softmax}(\frac{QK^T}{\sqrt{d_k}})V \enspace .
\end{align*}

\section{Prompts}
\label{app:prompts}

\RecustomVerbatimCommand{\VerbatimInput}{VerbatimInput}%
{fontsize=\footnotesize,
 frame=lines,  %
 framesep=2em, %
 rulecolor=\color{Gray},
 label=\fbox{\color{Black}indistribution.txt},
 labelposition=topline,
 commandchars=\|\(\), %
 commentchar=*        %
}

\begin{figure}[t]
\begin{Verbatim}
Translate the following natural language sentences into an LTL formula
and explain your translation step by step. Remember that X means
"next", U means "until", G means "globally", F means "finally",
which means GF means "infinitely often". The formula should only
contain atomic propositions or operators |, &, ~, ->, <->, X, U, G, F.
Natural Language: Every a is eventually followed by a e.
Given translations: {}
Explanation: "Every" from the input sentence refers to the temporal
operator "G", meaning that the subsequent part of the input must
hold at every point in time, i.e., globally. "eventually" from the
input sentence translates to the temporal operator "F".
"a followed by a e" means that after "a" holds, "e" has to hold as
well, i.e., translating to an implication. Thus, "a eventually
followed by a e" translates to "a -> F e". Explanation dictionary:
{"Every": "G", "eventually": "F", "a": "a", "e": "e", "a followed
by a e": "a -> e", "a eventually followed by a e": "a -> F e"}
So the final LTL translation is G (a -> F e).FINISH
Natural Language: a and b never occur at the same time but one of
them holds in every time step. Given translations: {}
Explanation: "a and b" from the input translates to the
conjunction of atomic propositions a,b, i.e., it
translates to "a & b". "a and b never occur" from the input
translates to the temporal behavior that at all positions, i.e.,
globally, neither a nor b hold, i.e., "G(~(a & b))". The input
additionally requires that "one of them holds in every time step",
which means that a or b hold globally, i.e., it translates
to "G(a | b)". Explanation dictionary: {"a and b": "a & b",
"a and b never occur": "G(~(a & b))", "one of them holds in every
time step": "G(a | b)"} So the final LTL translation is
G(~(a & b)) & G(a | b).FINISH Natural language: a can only
happen if b happend before. Given translations: {} Explanation:
"if b happend before" from the input means that until some point
b will happen, i.e., it translates to "U b" and "a can only
happen", means that a is not allowed to hold, i.e., it
translates to "~ a". In combination, this sentence represents that
something should not happen until a certain event happens, in this
case a is not allowed to hold until b holds. Explanation dictionary:
{"if b happend before": "U b", "a can only happen": "~ a"} So the
final LTL translation is (~ a) U b.FINISH Natural language: a holds
until b holds or always a holds. Given translations: {} Explanation:
"a holds until b holds" from the input translates to the temporal
modality "a U b" and "or" from the input translates to the
operator "|". "always a holds" from the input means that globally
a holds, i.e., it translates to "G a". Explanation dictionary:
{"a holds until b holds": "a U b", "or": "|", "always a holds":
"G a"} So the final LTL translation is (a U b) | G a.FINISH
\end{Verbatim}
\caption{Prompt engineered after the user study has been conducted, by drawing four random examples from the data set.}
\label{fig:in_distribution_prompt}
\end{figure}

Figure~\ref{fig:in_distribution_prompt} shows the prompt that has been augmented after the expert user study has been conducted.
The initial tutorial of LTL (line $1-5$) has been intentionally left the same as in the minimal prompt, to show the impact of the sub-translations.
The prompting leaves a lot of room for optimization, which is, however, out of the scope of this paper.
For example the combination of temporal operators \texttt{FG} is not explained, leading to failed translations in the expert user data set.

As a proof of concept, we provide two additional generic examples of constructing a prompt.
One is based on a recent case study on a smart home application~\cite{jarvis} (see Figure~\ref{fig:smart}), the second is a prompt Signal Temporal Logic (STL)~\cite{maler2004monitoring}, which is a variation of LTL on continuous signals (see Figure~\ref{fig:stl}).
Both prompts will be included in the corresponding open-source release of the tool.

\RecustomVerbatimCommand{\VerbatimInput}{VerbatimInput}%
{fontsize=\scriptsize,
 frame=lines,  %
 framesep=2em, %
 rulecolor=\color{Gray},
 label=\fbox{\color{Black}stl.txt},
 labelposition=topline,
 commandchars=\|\(\), %
 commentchar=*        %
}

\begin{figure}[t]
\begin{Verbatim}
Translate the following natural langauge sentence into an STL formula and explain your
translation step by step. Let a and b be variables. Remember that U[a,b] means
"until the interval a and b", "F [a,b]" means eventually in the interval between
a and b, and always is "G[a,b]". Additionally, STL consists of predicates.
Assume signals x1[t], x2[t], . . . , xn[t], then atomic predicates are of the form:
f(x1[t], . . . , xn[t]) > 0. The STL formula should only contain atomic propositions,
boolean operators &, ~, ->, <-> and temporal operators U[a,b], G[a,b], F[a,b].
Natural Language: The signal is never above 3.5. Given translations: {}
Explanation: "The signal" from the input translates to the variable "x[t]", "above 3.5"
from the input translates to "> 3.5", "never above 3.5" thus means that the signal
should never be above 3.5, i.e., always under 3.5. Explanation dictionary:
{"The signal": "x[t]", "above 3.5": "> 3.5", "never above 3.5":
"G < 3.5"} So the final STL translation is G (x[t] < 3.5).
Natural Langauge: Between 2s and 6s the signal is between -2 and 2. Given
translations: {} Explanation: "Between 2s and 6s" from the input translates to the
temporal operator "G[2,6]" and "the signal is between -2 and 2" translates to the
predicate "abs(x[t]) < 2". Explanation dictionary: {"Between 2s and 6s": "G[2,6]",
"the signal is between -2 and 2": " abs(x[t]) < 2"}.
So the final STL translation is G[2,6] (abs(x[t]) < 2).
\end{Verbatim}
\caption{Example prompt for Signal Temporal Logic (STL). Few-shot examples are taken from a talk~\cite{donze2013signal}.}
\label{fig:stl}
\end{figure}

\RecustomVerbatimCommand{\VerbatimInput}{VerbatimInput}%
{fontsize=\scriptsize,
 frame=lines,  %
 framesep=2em, %
 rulecolor=\color{Gray},
 label=\fbox{\color{Black}smart.txt},
 labelposition=topline,
 commandchars=\|\(\), %
 commentchar=*        %
}

\begin{figure}[t]
\begin{Verbatim}
Translate the following natural language sentences into an LTL formula and explain your
translation step by step. Remember that X means "next", U means "until", G means
"globally", F means "finally", which means GF means "infinitely often". The formula
should only contain atomic propositions or operators |, &, ~, ->, <->, X, U, G, F.
Natural language: The coffee machine is always ready when somebody is at the room.
Given translations: {} Explanation: "The coffee machine is ready" from the input
translates to "c" and "always" translates to "G". "Somebody is at the room" from the
input translates to "r". Explanation dictionary: {"The coffee machine is ready" :
"c", "always" : "G", "somebody is at the room" : "r"} So the final LTL translation
is G (r -> c).FINISH Natural language: Lights are only on if somebody is in the room.
Given translations: {} Explanation: "Lights are on" from the input translates to
"l" and "somebody is in the room" translates to "r". "only if" from the input
translates to an equivalence "<->". Additionally, there is an implicit meaning
that this is always the case, which translates to "G". Explanation dictionary:
{"Lights are on" : "l", "somebody is in the room" : "r", "only if" : "<->"}
So the final LTL translation is G (l <-> r).FINISH
\end{Verbatim}
\caption{Example prompt of two few-shot examples from a smart home case study~\cite{jarvis}.}
\label{fig:smart}
\end{figure}

\section{Remaining Instances}
\label{app:remaining_instances}
In the following, we report the instances that we were unable to translate with \texttt{nl2spec} and the minimal prompt in $\leq 3$ translation tries.
These instances are especially difficult since their writers use advanced operators like ``release'' or hid the temporal meaning inside the natural language:

{\centering
\resizebox{\linewidth}{!}{
\begin{tabular}{c | c}
    natural language S & LTL specification $\psi$ \\
    \midrule
    Once a happened, b won't happen again. & \texttt{G (a -> X G ! b)} \\
    \midrule
    a releases b & \texttt{(b U (b \& ! a)) | G b} \\
    \midrule
    a holds in every fifth step. & \texttt{a \& G (a -> X ! a \& X X ! a}\\
    ~& \texttt{\& X X X ! a \& X X X X ! a \& X X X X X a)} \\
    \midrule
    a must always hold, but if is execeeds, & \texttt{! G (! (a \& X a))} \\
    it allow two timestamps to recover. & ~\\
    \midrule
    not a holds at most two timestamps & \texttt{! G (! (a \& X a))} \\
    \bottomrule
\end{tabular}
}
}

\section{Variables}
\label{app:variables}

\begin{figure}[h]
    \centering
    \includegraphics[width=\textwidth,height=0.5\textheight]{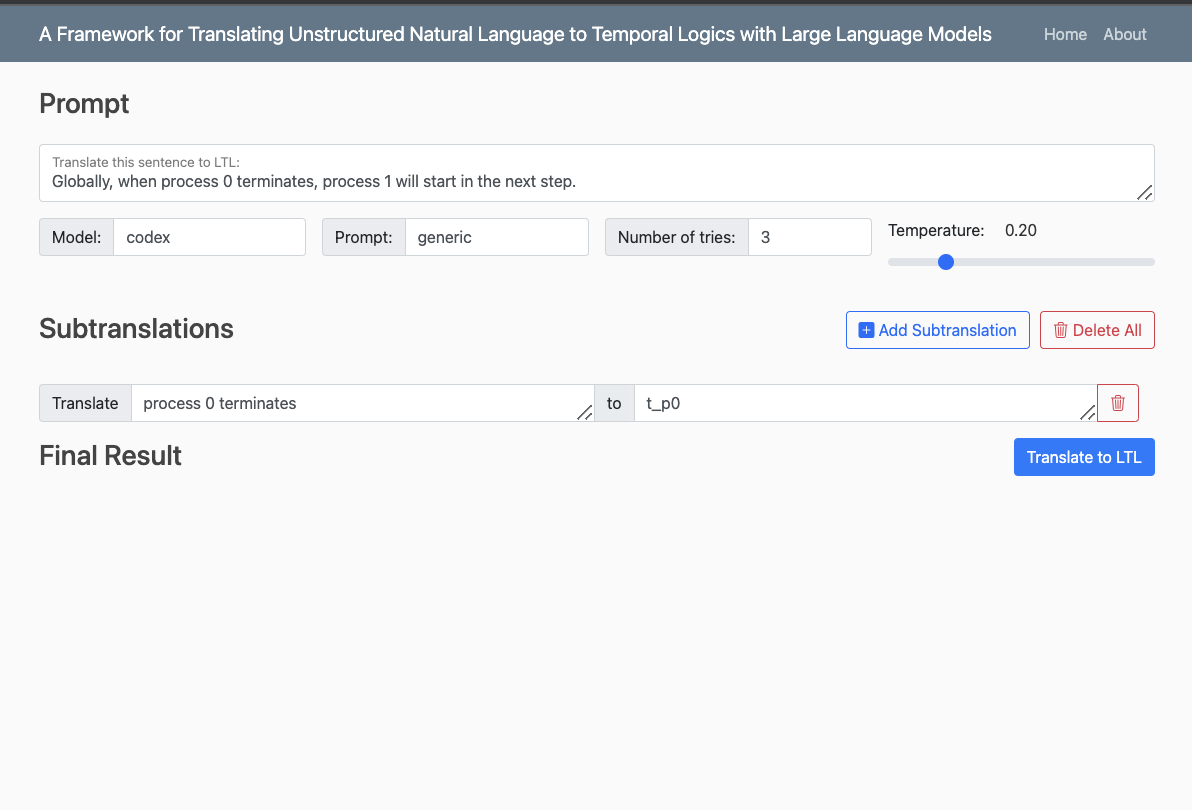}
    \includegraphics[width=\textwidth,height=0.5\textheight]{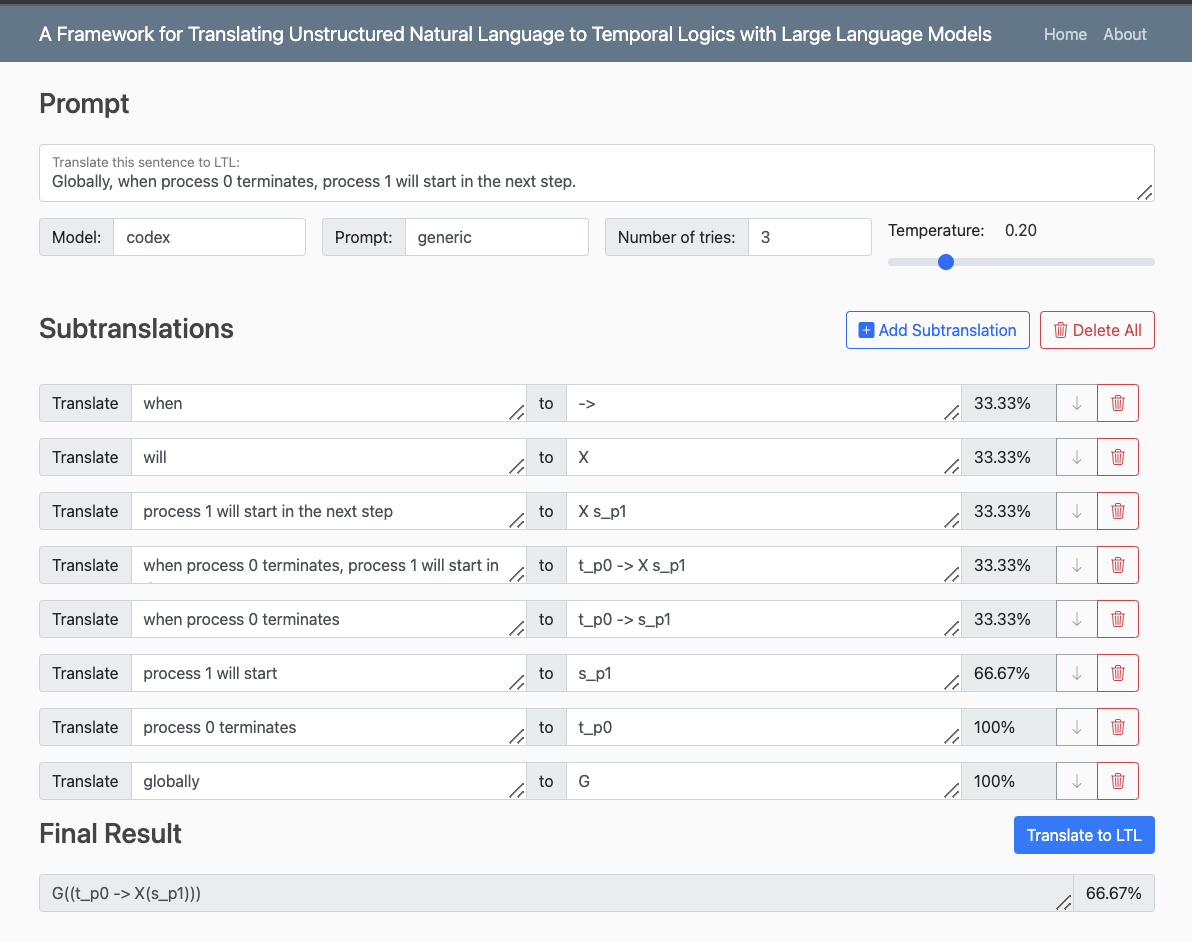}
    \caption{Providing a sub-translation for variable translations.}
    \label{fig:variable}
\end{figure}

An advantage of using large language models is their ability to adjust to patterns in natural language.
Translations of variables can just be given or edited as subtranslations.
For example, consider the following natural language input ``Globally, when process 0 terminates, process 1 will start in the next step''.
By adding the sub-translation ``process 0 terminates'' as \texttt{t\_p0} and startin the translation, the model adjusts the variable for ``process 1 will start'' automatically as \texttt{s\_p1} (see Figure~\ref{fig:variable}).

\section{Ambiguity Resolving}
\label{app:ambiguity_resolving}

\begin{figure}[h]
    \centering
    \includegraphics[width=\textwidth,height=0.38\textheight]{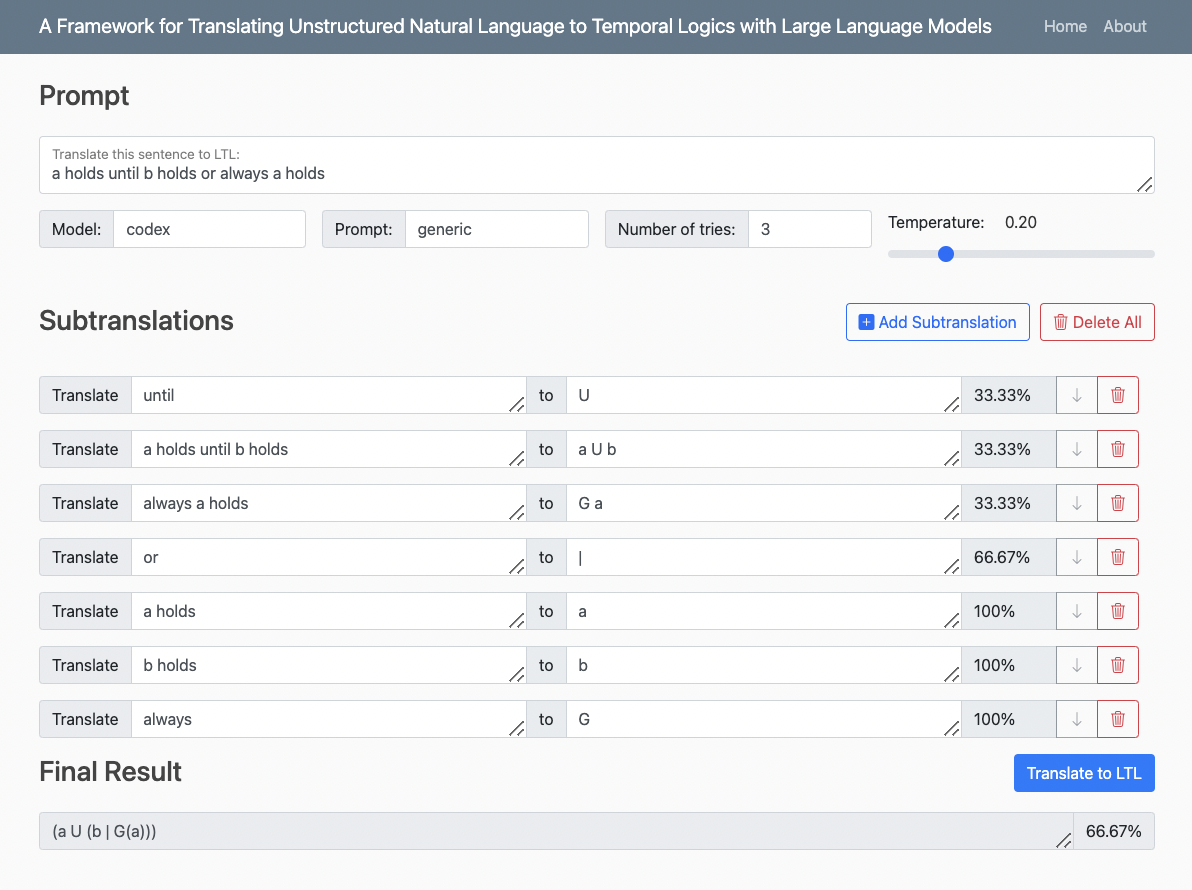}
    \includegraphics[width=\textwidth,height=0.38\textheight]{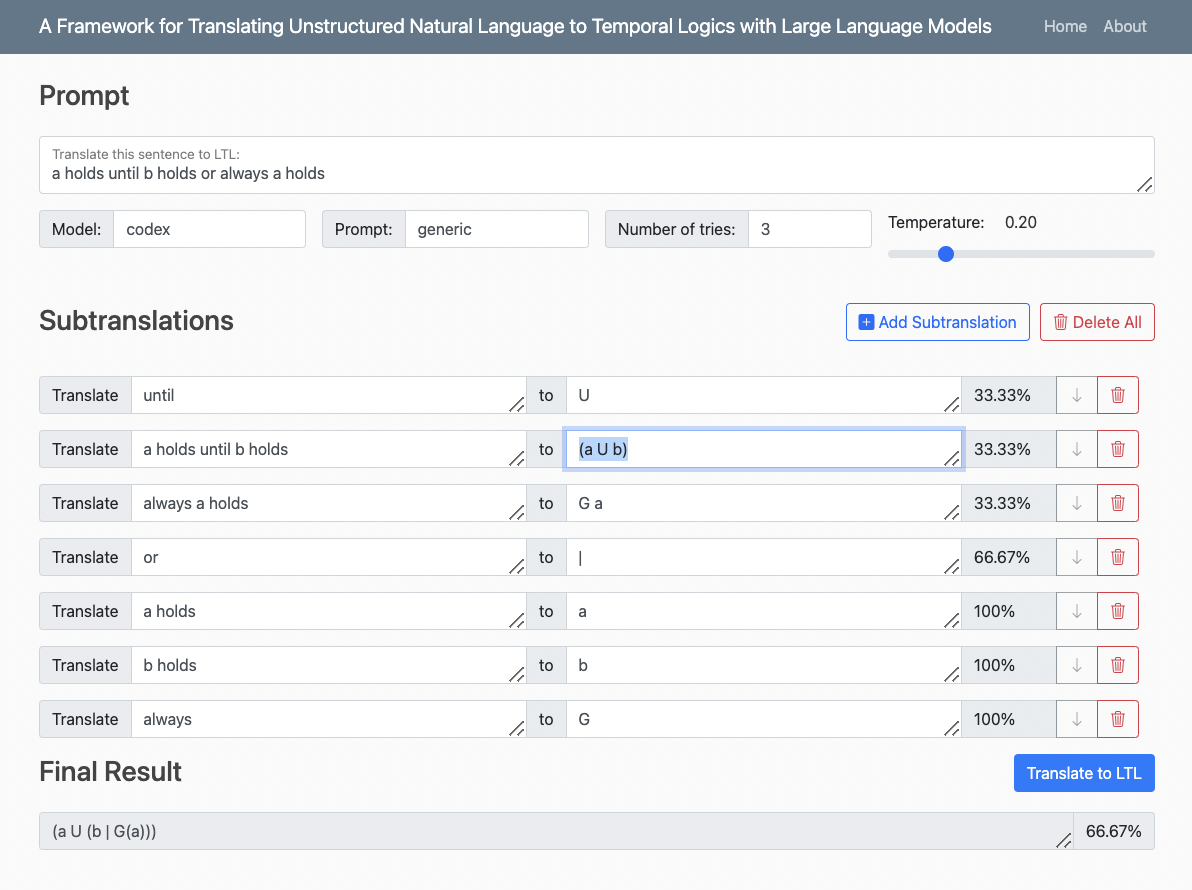}
    \includegraphics[width=\textwidth,height=0.38\textheight]{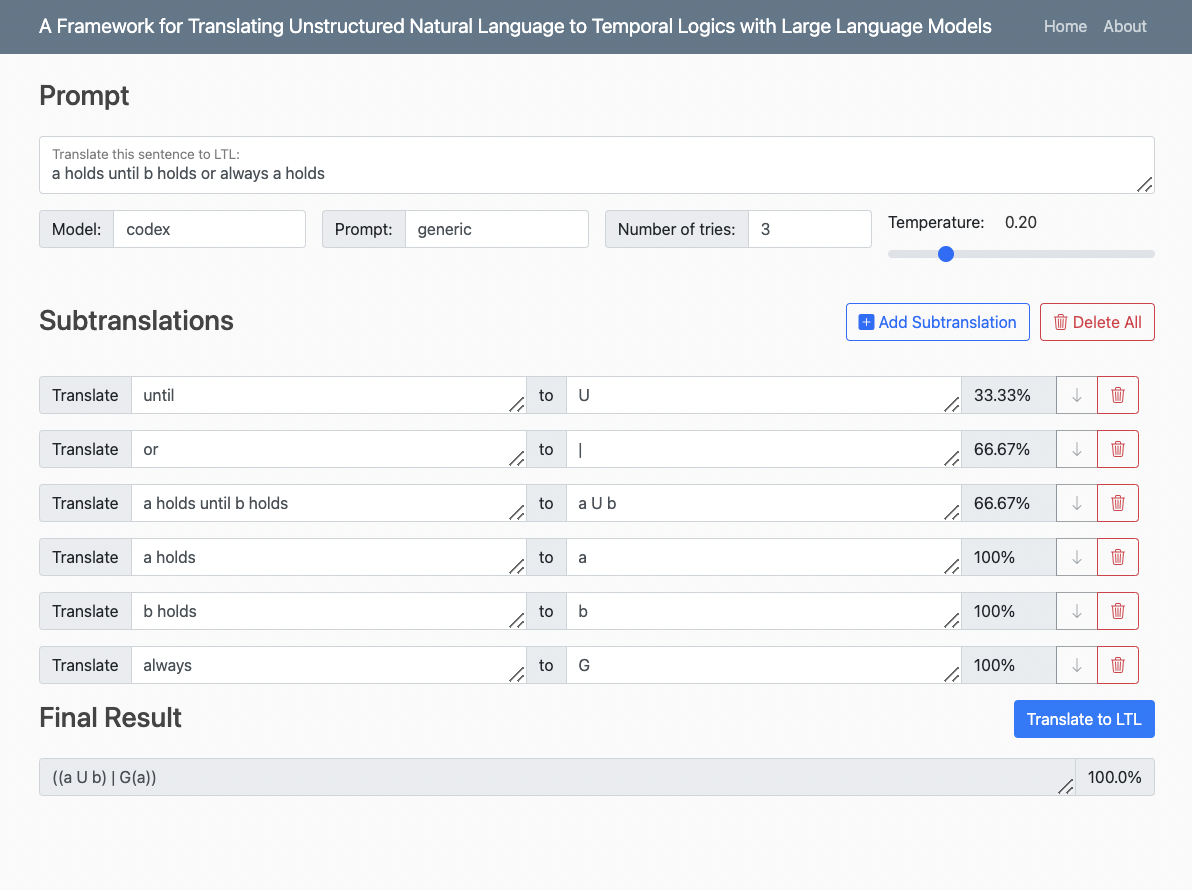}
    \caption{Fixing the subtranslation with parenthesis, to ensure operator precedence.}
    \label{fig:amb_3}
\end{figure}

\begin{figure}[h]
    \centering
    \includegraphics[width=\textwidth,height=0.38\textheight]{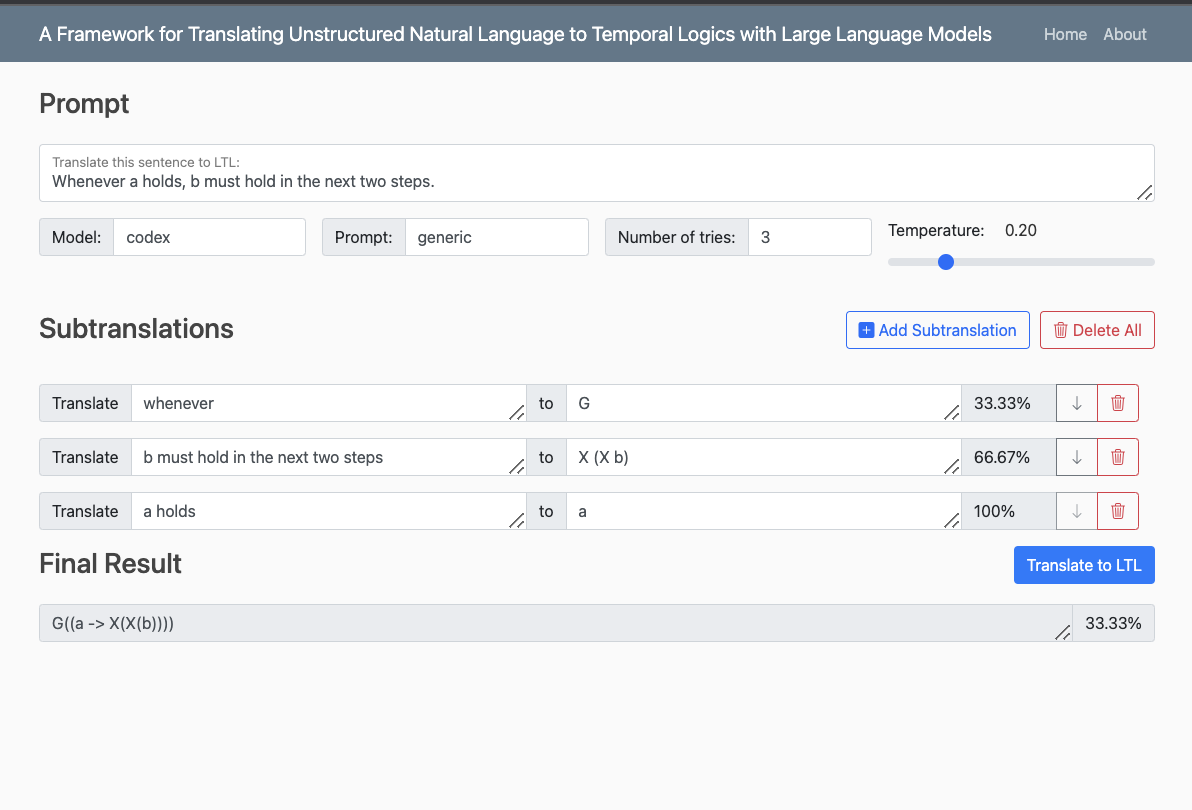}
    \includegraphics[width=\textwidth,height=0.38\textheight]{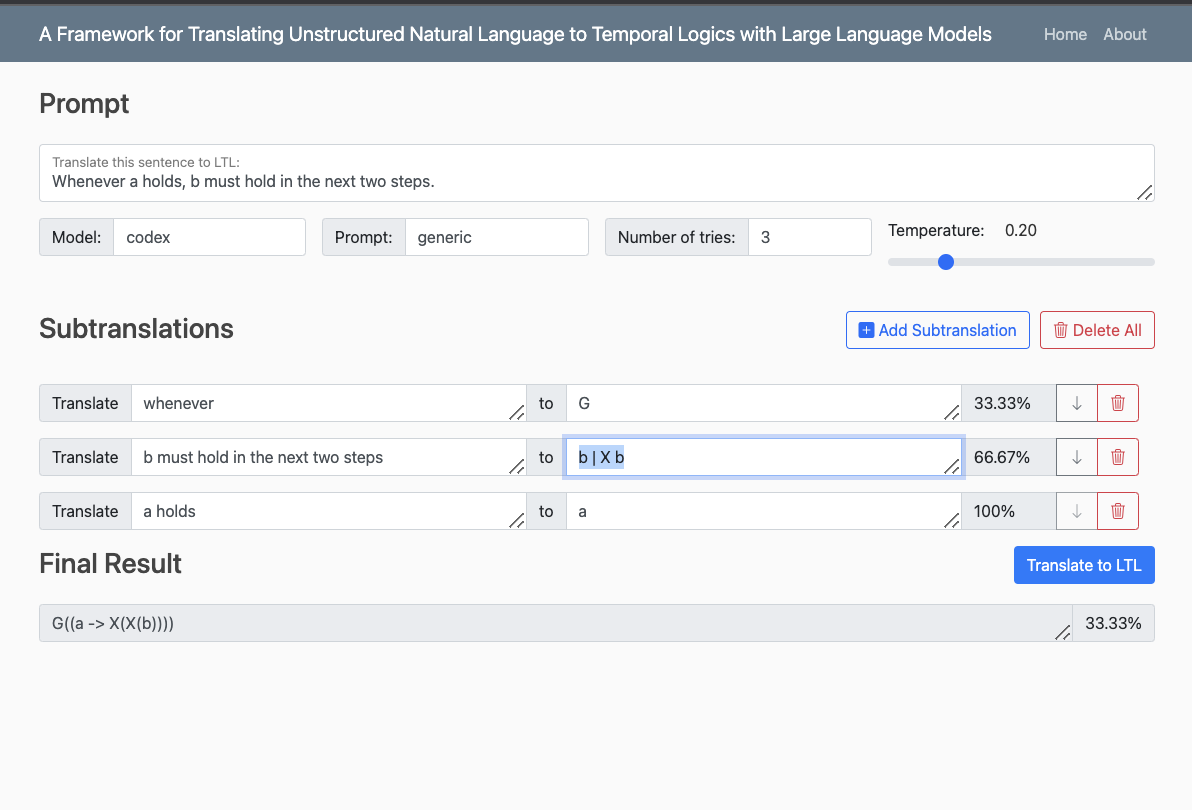}
    \includegraphics[width=\textwidth,height=0.38\textheight]{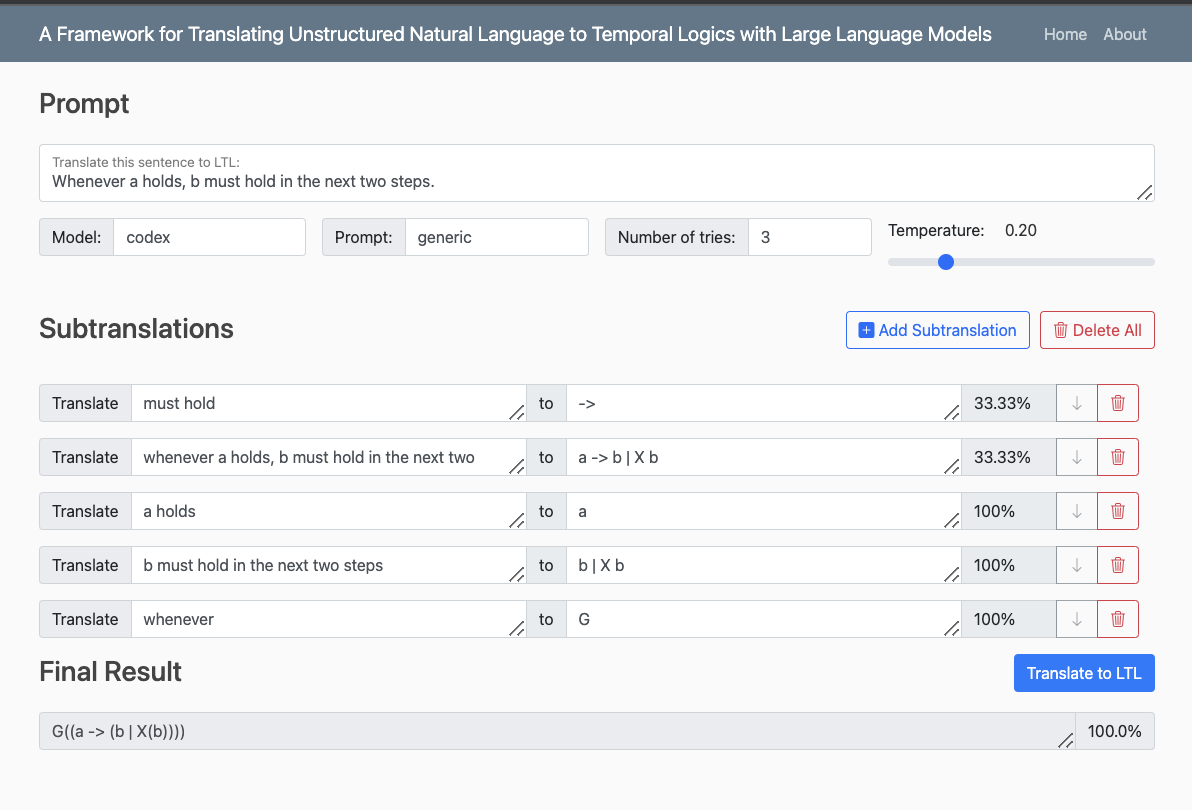}
    \caption{Editing a subtranslation to reflect the meaning of the input.}
    \label{fig:amb_second_example}
\end{figure}

Figure~\ref{fig:amb_3} and Figure~\ref{fig:amb_second_example} show screenshots of the frontend of \texttt{nl2spec} while resolving the ambiguity of the examples.

\section{Fixing Erroneous Translations}
\label{app:fixing}
\begin{figure}[h]
    \centering
    \includegraphics[width=\textwidth,height=0.38\textheight]{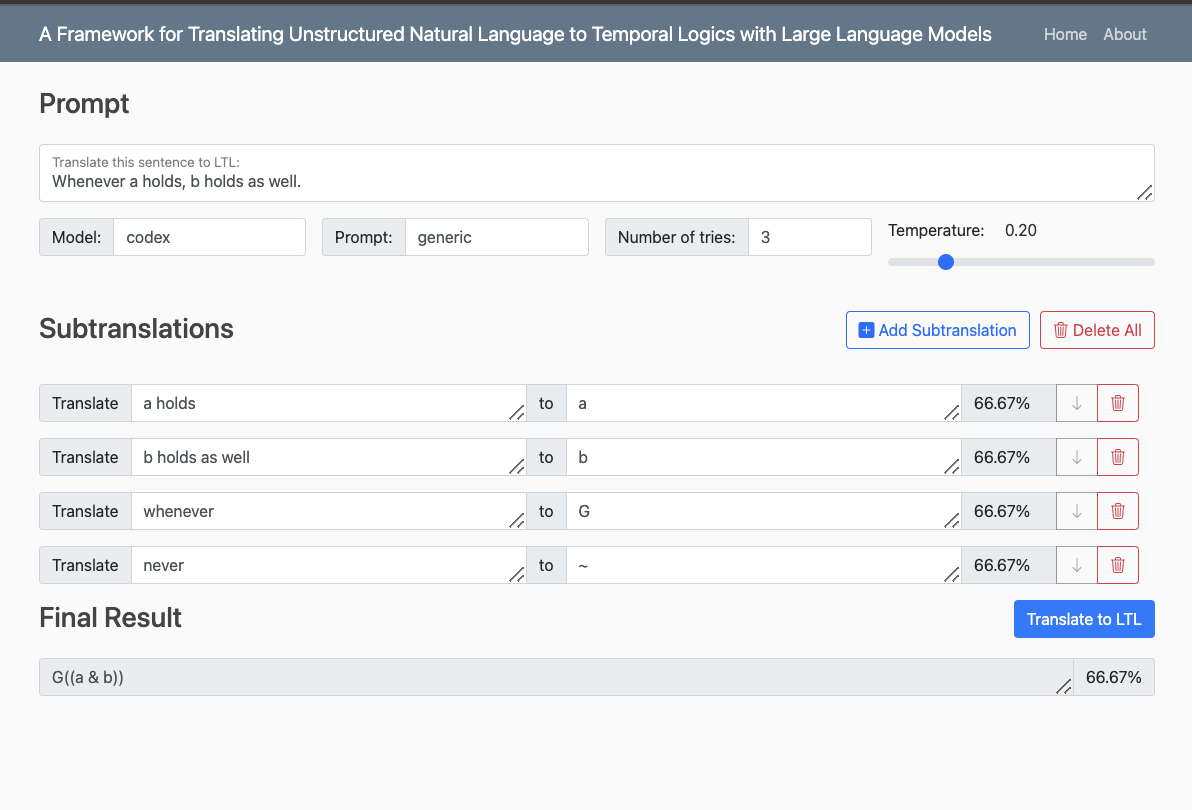}
    \includegraphics[width=\textwidth,height=0.38\textheight]{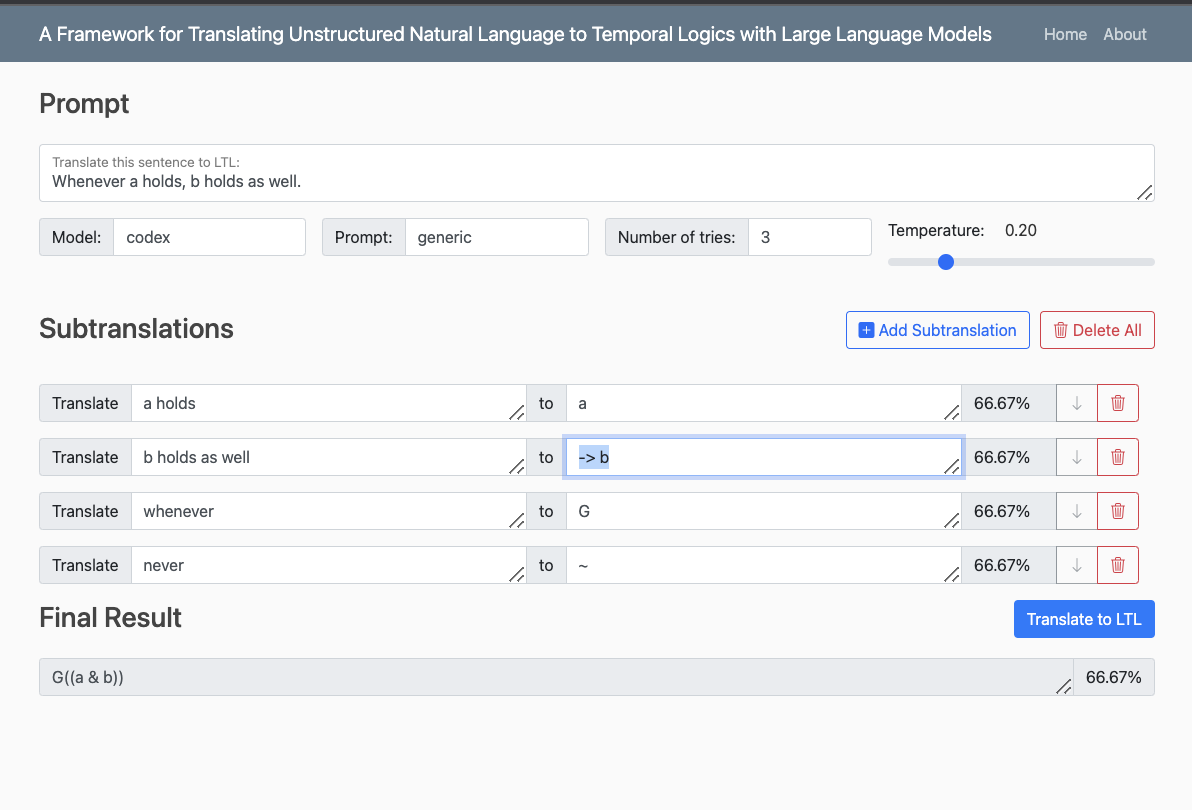}
    \includegraphics[width=\textwidth,height=0.38\textheight]{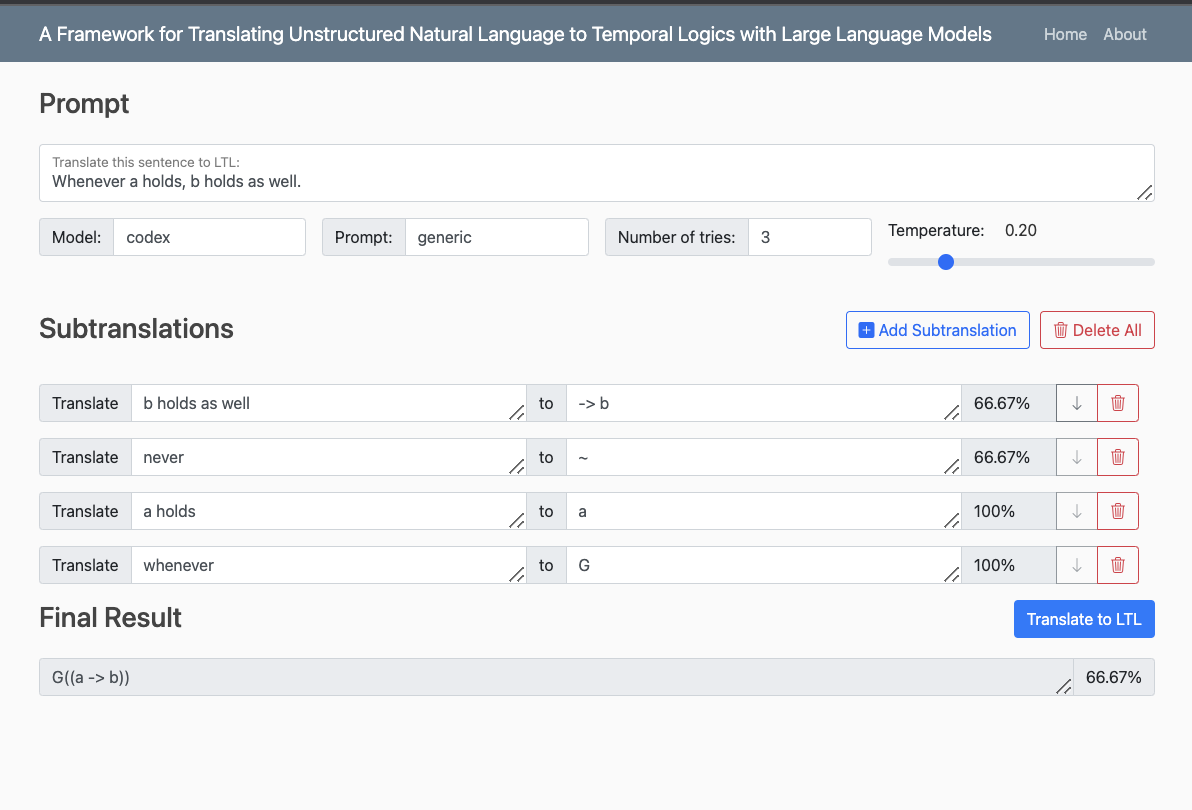}
    \caption{Editing a subtranslation to reflect the meaning of the input.}
    \label{fig:edit}
\end{figure}

Figure~\ref{fig:edit} shows a screenshots of the frontend of \texttt{nl2spec} when editing a sub-translation to debug an erroneous translation.

\end{document}